\documentclass[pre,twocolumn,showpacs,aps,amsmath,amssymb,floatfix,superscriptaddress]{revtex4}
\pdfoutput=1 %This is for the ArXiv submission only
\usepackage{graphicx,eucal,psfrag}
\usepackage{epsfig}

\newcommand{\ra}{\rangle}
\newcommand{\la}{\langle}
\newcommand{\be}{\begin{equation}}
\newcommand{\ee}{\end{equation}}

\begin{document}
\title{Compact Waves in Microscopic Nonlinear Diffusion}

\author{P. I. Hurtado}
\affiliation{Institute Carlos I for Theoretical and Computational Physics, and Departamento de Electromagnetismo y F\'{\i}sica de la Materia, Universidad de Granada, 18071 Granada, Spain}

\author{P. L. Krapivsky}
\affiliation{Department of Physics, Boston University, Boston, Massachusetts 02215, USA}

\begin{abstract}
We analyze the spread of a localized peak of energy into vacuum for nonlinear diffusive processes. In contrast with standard diffusion, the nonlinearity results in a compact wave with a sharp front separating the perturbed region from vacuum. In $d$ spatial dimensions, the front advances as $t^{1/(2+da)}$ according to hydrodynamics, with $a$ the nonlinearity exponent. We show that fluctuations in the front position grow as $\sim t^{\mu}\eta$, where $\mu<\tfrac{1}{2+da}$ is a new exponent that we measure and $\eta$ is a random variable whose distribution we characterize. Fluctuating corrections to hydrodynamic profiles give rise to an excess penetration into vacuum, revealing scaling behaviors and robust features. We also examine the discharge of a nonlinear rarefaction wave into vacuum. Our results suggest the existence of universal scaling behaviors at the fluctuating level in nonlinear diffusion.
\end{abstract}
\maketitle

Random walks, Brownian motions, and their continuum descriptions in terms of diffusion equations underlie uncountable natural phenomena \cite{B93,V01}. Our experience with diffusion strongly influences our intuition. For instance, a peculiar property of the diffusion equation, viz. the instantaneous propagation of perturbations, appears puzzling when we first learn it, but eventually we regard it as a general property of parabolic partial differential equations (PDEs) which distinguishes them from hyperbolic PDEs. It then comes as a surprise that the instantaneous propagation of perturbations is lost for certain parabolic PDEs, so that phenomenologically they start to resemble hyperbolic PDEs. A simple class of parabolic PDEs which exhibits such a property is 
\begin{equation}
\label{NLDE:general}
\partial_t \rho = \nabla\cdot \left(\rho^a\,\nabla \rho\right)
\end{equation}
For such equations, the spread of a initially localized profile into vacuum is \emph{not} instantaneous whenever $a>0$. More precisely, in $d$ dimensions the front advances as $r_f\sim t^{1/(2+da)}$; for $r>r_f$, the medium is still in the vacuum state \cite{Zeld,LL87,B96,book}. There is no contradiction with standard lore as Eq.~\eqref{NLDE:general} is {\em non-linear}. Further, this spectacular effect occurs only when we consider the spread into vacuum (where diffusion coefficient $D=\rho^a$ vanishes). 

Nonlinear parabolic PDEs form a fertile research area. We emphasize that  nonlinear parabolic PDEs similar to Eq.~\eqref{NLDE:general} and their microscopic brethren are by no means pathological. For instance, Navier-Stokes equations are nonlinear parabolic PDEs, and e.g. in heat conduction the coefficient of thermal conductivity always depends on temperature \cite{Zeld,LL87}, for instance it scales as $\sqrt{T}$ for the hard-sphere gas indicating that the mathematical description is similar to Eq.~\eqref{NLDE:general} with $a=\frac{1}{2}$. 

Here we investigate microscopic stochastic processes which admit a macroscopic description in terms of Eq.~\eqref{NLDE:general}. Amending microscopic rules, it is usually possible to vary the exponent $a$. For concreteness, we focus on a specific exponent $a=1$. More importantly, we consider only simple microscopic models as our major goal is to go beyond hydrodynamics and to analyze fluctuations. 

Two microscopic processes that meet above requirements are the Averaging Process (AP) and the Random Exchange Process (REP). These are continuous-time Markov models that can be defined on an arbitrary lattice. At a coarse-grained level these two models mimic the physics of systems which are characterized by a single locally-conserved field. By definition, each site $i$ carries a scalar variable $\rho_i\geq 0$ which we call energy. For concreteness, we start with the one-dimensional (1d) lattice. For the AP, the evolution proceeds by choosing a bond at random, say $(i,i+1)$, and averaging the energies of the sites connected by the bond, $(\rho_i,\rho_{i+1})\to (\frac{1}{2}\Sigma_i,\frac{1}{2}\Sigma_i)$, with $\Sigma_i\equiv \rho_i+\rho_{i+1}$. For the REP, the pair energies are randomly redistributed, $(\rho_i,\rho_{i+1})\to (p\Sigma_i,(1-p)\Sigma_i)$, where $p\in[0,1]$ is a random number chosen uniformly on the interval $[0,1]$ in each ``collision'' event. If collisions occur independently on the energies, the above stochastic processes are one of the simplest microscopic implementations of the diffusion process \cite{kmp}, while energy-dependent collision rates yield nonlinear diffusion. 
%By writing the evolution equation for the different moments of the local energy and closing the subsequent hierarchy by assuming that higher-order averages factorize \cite{plh}, we find that the macroscopic evolution equation for the AP is \cite{longpaper}

Consider for concreteness the AP. If the collision rate is proportional to the total pair energy $\Sigma_i$, the energy $\rho_i(t+dt)$ at site $i$ after an infinitesimal time increment $dt$ will become $\frac{1}{2}\Sigma_{i-1}$ with probability $\Sigma_{i-1}dt$, $\frac{1}{2}\Sigma_i$ with probability $\Sigma_{i}dt$, or will remain as $\rho_i(t)$ with probability $1-(\Sigma_{i-1}+\Sigma_i)dt$. Averaging and taking the limit $dt\to 0$ we obtain
\begin{equation}
\label{E_av}
\frac{d\la\rho_i\ra}{dt}=\frac{1}{2}\Big(\la \rho_{i-1}^2\ra - 2\la \rho_i^2\ra + \la \rho_{i+1}^2\ra\Big) \, ,
\end{equation}
where we write in short $\la\rho_i\ra\equiv \la\rho_i(t)\ra$, $\la\rho_i^2\ra\equiv \la[\rho_i(t)]^2\ra,$ etc. The governing equation (\ref{E_av}) for average energies involves second-order averages, whose time evolution is controlled in turn by  higher-order averages, giving rise to an infinite hierarchy which is not tractable in general. This hierarchy can be closed however by assuming that higher-order averages factorize \cite{plh}, i.e. $\la\rho_i^2\ra=\la\rho_i\ra^2$. This conjecture, a central tenet of the hydrodynamic approach, is expected to become asymptotically valid in the long time limit \cite{book}, where evolution is dominated by the slowly-varying, locally-conserved hydrodynamic field.
%We can however use a hydrodynamic, or mean-field, approach based on the conjecture that higher-order averages factorize. We thus assume that $\la\rho_i^2\ra=\la\rho_i\ra^2$, an ansatz which can become asymptotically valid only in the long time limit \cite{book}. 
In this situation the spatial variance is very small, so that we can replace the discrete (second-order) derivative on the right-hand side of \eqref{E_av} by a continuous derivative. Changing now the spatial variable $i\to x$ to emphasize that we are employing the continuum approximation, and defining $\rho(x,t)\equiv \la\rho_x(t)\ra$, we arrive at
\be
\partial_t\rho=\partial_x(\rho\, \partial_x\rho) \, ,
\label{NLDE}
\ee
i.e. Eq.~\eqref{NLDE:general} with exponent $a=1$. [For the REP we similarly obtain $\partial_t\rho=2\partial_x(\rho\, \partial_x\rho)$.] Generally the AP and REP processes with collision rates proportional to a power of the pair energy, $\Sigma_i^a$, are the examples  \cite{longpaper,plh} of microscopic stochastic processes whose hydrodynamics is governed by the generalized nonlinear diffusion equation \eqref{NLDE:general}. Needless to say, fluctuating behaviors are outside the realm of the hydrodynamic approach.

First, we consider the evolution of a localized peak of energy: $\rho(x,t=0)=\rho_0 \, \delta(x)$, where $\rho_0 = \int_{-\infty}^\infty dx\, \rho(x,t)$ is the total energy which remains constant. The lack of characteristic spatial and temporal scales in our initial-value problem suggests to apply the transformation $x\to \alpha x, ~t\to \beta t, ~\rho\to \gamma \rho$. Equation \eqref{NLDE} is invariant under this transformation when $\alpha^2=\beta\gamma$, and the conservation of energy leads to another constraint $\alpha \gamma = 1$. Therefore $\alpha=\gamma^{-1}=\beta^{1/3}$, and our problem remains invariant under the one-parameter transformation group $x\to \beta^{1/3} x, ~ t\to \beta t, ~ \rho\to \beta^{-1/3} \rho$, which suggests to seek a solution in a self-similar form
\be
\label{scaling}
\rho(x,t) = \frac{\rho_0}{(\rho_0 t)^{1/3}}\,\Phi(\xi), \quad \xi = \frac{x}{(\rho_0 t)^{1/3}}
\ee
Plugging ansatz \eqref{scaling} into the governing equation and integrating once we find that the scaled energy profile $\Phi$ satisfies $\Phi(\Phi'+\tfrac{1}{3}\xi)= 0$, where prime denotes the differentiation with respect to $\xi$. This equation reduces to $\Phi' = -\xi/3$ in a bump region $|\xi| \leq \xi_0$ and to $\Phi=0$ outside the bump:
\be
\label{parabola}
\Phi = 
\begin{cases}
A(\xi_0^2-\xi^2) & |\xi| \leq \xi_0\\
0                                      &|\xi| > \xi_0
\end{cases}
\ee
with $A=1/6$. The conservation of energy $\int d\xi\,\Phi(\xi)=1$ fixes the boundaries of the bump: $\xi_0=(9/2)^{1/3}$. (For the REP, due to the extra factor 2 in its hydrodynamic equation, $A=1/12$ and $\xi_0=9^{1/3}$.) The most striking feature of the (parabolic) energy profile \eqref{parabola} is that it is compact --- the energy density vanishes at a finite distance from the initial peak. This contrasts with standard expectations based on the behavior of solutions to the {\em linear} diffusion equation where the influence propagates instantaneously --- when diffusion is non-linear, the `knowledge' about the peak at the origin is accessible only on distances $|x|\leq \xi_0\left(\rho_0 t\right)^{1/3}$.

The self-similar solution \eqref{scaling}--\eqref{parabola} was derived in \cite{ZK,B52};  similar solutions were found earlier in studies of the intermediate stage of nuclear explosions. These solutions have been investigated in the realm of hydrodynamics, which says nothing about fluctuations. In what follows we analyze fluctuating corrections to the average hydrodynamic behavior using both the AP and REP stochastic models. In particular, we run extensive simulations of both models, averaging over many different realizations with fixed initial condition $\rho_i(t=0)=\rho_0 \delta_{i,0}$.

\begin{figure}
%\vspace{-0.5cm}
\centerline{
%\vspace{-0.5cm}
\includegraphics[width=8cm]{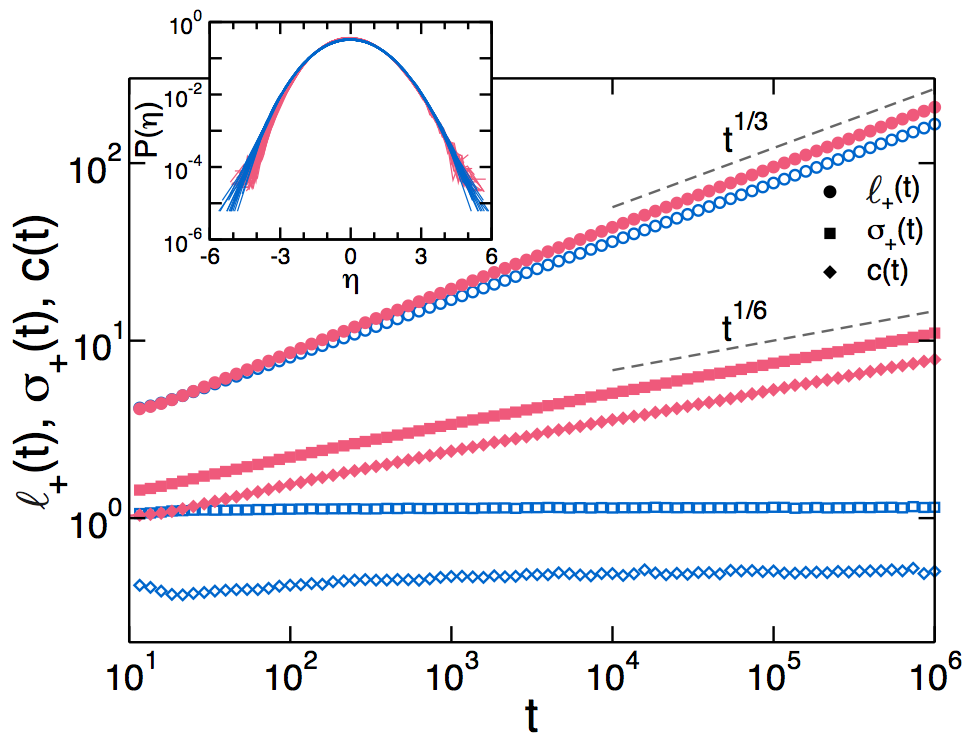}}
\caption{(Color online) Right front position ($\bigcirc$), its standard deviation ($\Box$) and correlation between opposite fronts ($\Diamond$) measured for the AP (blue, open) and the REP (red, filled) models. Inset: Distribution of front fluctuations for the AP and the REP.}
\label{front}
\end{figure}

%\begin{figure}
%\psfrag{l}[b][b][1.2]{$\ell$}
%\vspace{-0.5cm}
%\centerline{
%\vspace{-0.5cm}
%\includegraphics[width=8cm]{leader-sigma-correl-averaging-and-kmp-v3.pdf}}
%\caption{(Color online) Right front position ($\bigcirc$), its standard deviation ($\Box$) and correlation between opposite fronts ($\Diamond$) measured for the AP (blue, open) and the REP (red, filled) models. Inset: Distribution of front fluctuations for the AP and the REP.}
%\label{front}
%\end{figure}

We first consider the right front position, represented at the microscopic level by the position $\ell_+(t)$ of the rightmost site where the energy at time $t$ is positive. The front position $\ell_+(t)$ is a random quantity. Conjecturally,  
\begin{equation}
\label{j+}
\ell_+(t) = \xi_0\left( \rho_0 t\right)^{1/3}  +  (\rho_0 t)^{\mu} \eta
\end{equation}
where $\mu<\frac{1}{3}$ is a new exponent and $\eta$ is a random variable. Figure \ref{front} shows the measured front position and its standard deviations $\sigma_+(t)$. This confirms that the front propagates as $t^{1/3}$ for both the AP and the REP. Interestingly, front fluctuations are characterized by an exponent $\mu=\frac{1}{6}$ for the REP model, while $\mu=0$ for the AP, i.e. much smaller fluctuations in the latter than in the former for long times. This can be attributed to the additional \emph{layer of stochasticity} (the random energy exchange) which characterizes the REP. We also measured the correlation between left and right fronts, $\kappa(t)\equiv [\la\ell_+\ell_-\ra - \la\ell_+\ra\la\ell_-\ra](t)$. Opposite fronts turn out to be anti-correlated, $\kappa(t)<0$, due to energy conservation, see Fig.~\ref{front} where we plot $c(t)\equiv \sqrt{-\kappa(t)}$. Moreover, we observe $c(t)\sim t^{\mu}$ with the same exponents $\mu$ as we measured above for the AP and the REP.

The probability distribution $\mathcal{P}(\eta)$ which characterizes front fluctuations is unknown, though its asymptotic behaviors can be understood heuristically. To estimate the $\eta\to -\infty$ tail let us first determine the probability that $\ell_+(t)=0$. Since $\ell_+$ can only increase, the event $\ell_+(t)=0$ implies that during the time interval $(0,t)$, the pair $(0,1)$ has never collided. 
The probability that this pair does not collide in an infinitesimal time interval $(t',t'+dt')$ is $[1-\rho(x=0,t')dt']\approx \exp[-\rho(0,t')dt']$, so 
$\ln \text{Prob}(\ell_+(t)=0)=-\int_0^t d\tau\, \rho(0,\tau)$.
%the probability  that this pair does not collide during a time t can be written as $\exp[-\int_0^t\rho(x=0,t')dt']$
Since we are not constraining the motion of the left front, the wave span still grows as $t^{1/3}$
%The span of the non-zero profile still scales as $t^{1/3}$, 
and hence the energy at origin decays as $\rho(0,t) \sim t^{-1/3}$ due to energy conservation, which implies that $\ln \text{Prob}(\ell_+(t)=0)\sim -t^{2/3}$.
%$\ln \text{Prob}(\ell_+(t)=0)=-\int_0^t d\tau\, \rho(x=0,\tau)\sim -t^{2/3}$.
%\begin{equation*}
%\ln \text{Prob}(\ell_+(t)=0)=-\int_0^t d\tau\, \rho(x=0,\tau)\sim -t^{2/3} \, .
%\end{equation*}
From \eqref{j+} we see that $\ell_+(t)=0$ corresponds to a fluctuation $\eta \sim -t^{(1-3\mu)/3}$. This in conjunction with above estimate yields $\ln \mathcal{P}(\eta) \sim - |\eta|^{2/(1-3\mu)}$ when $\eta\to -\infty$.

\begin{figure}
\vspace{-0.75cm}
\centerline{
\vspace{-0.15cm}
\includegraphics[width=8cm]{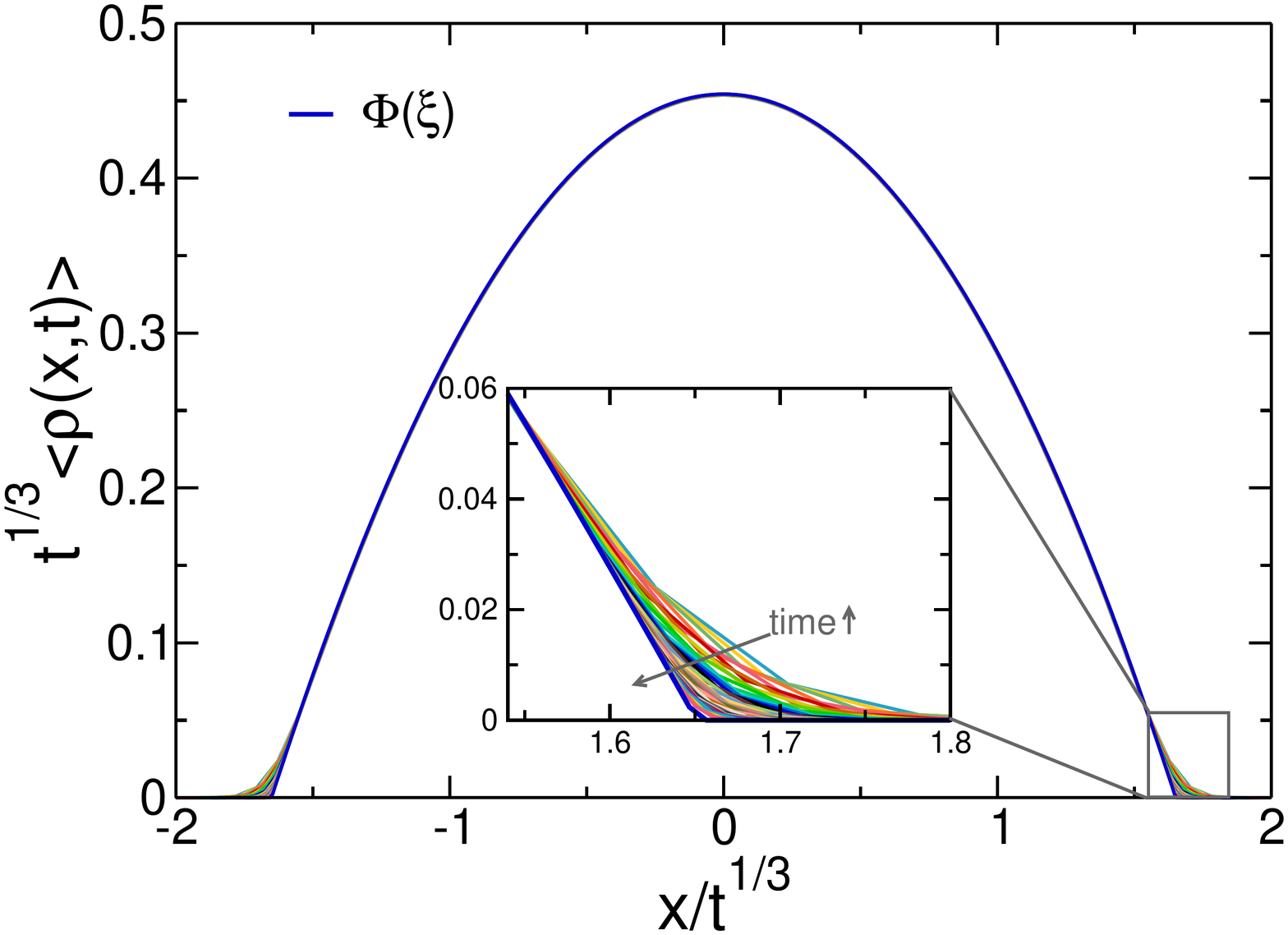}}
\centerline{
\vspace{-0.25cm}
\includegraphics[width=8cm,trim= 0mm 0mm 0mm 15mm,clip]{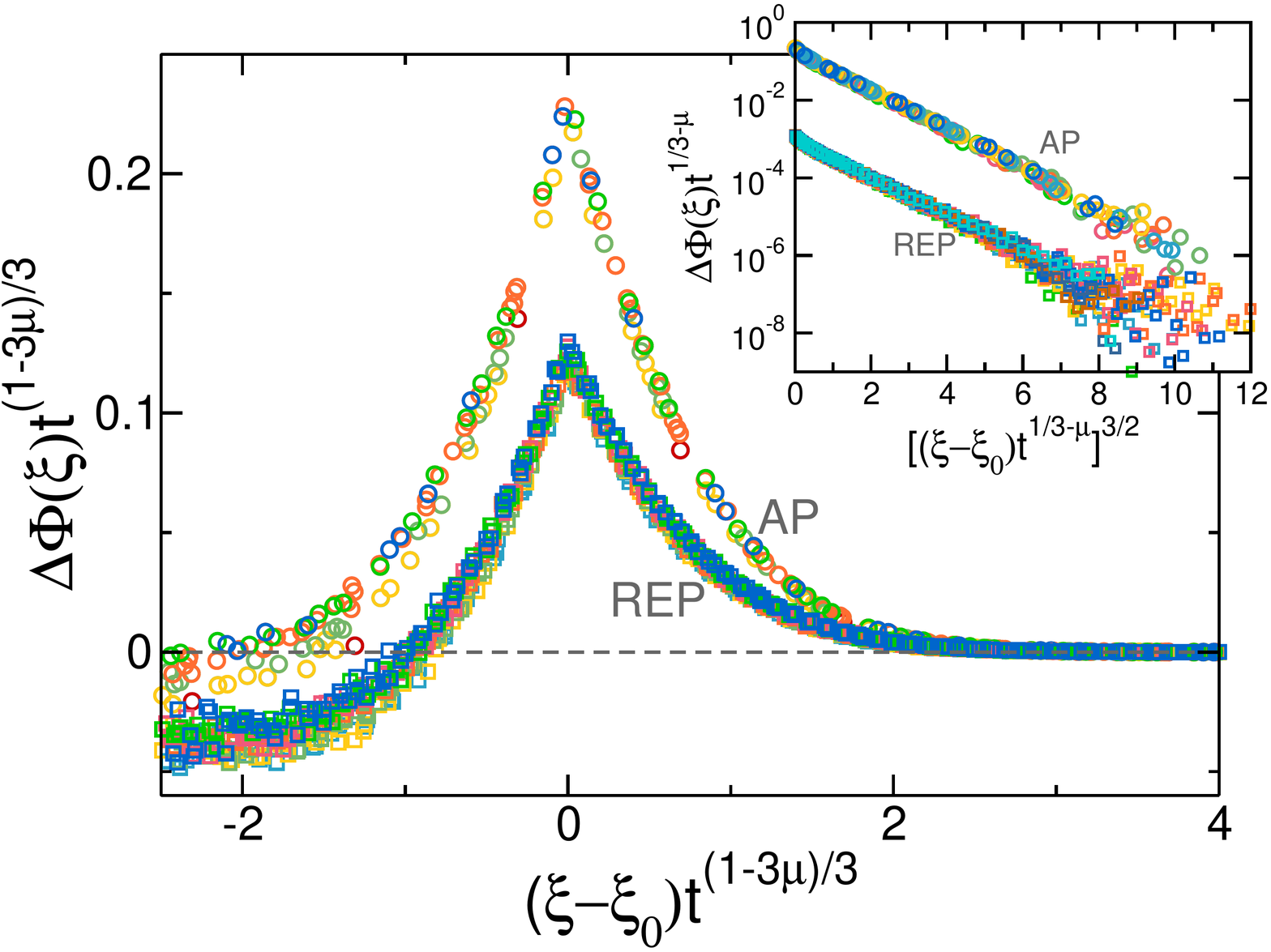}}
\caption{(Color online) Top: Scaled energy profiles for different times in the AP model and the theoretical prediction $\Phi(\xi)$. The inset shows the microscopic penetration into vacuum. Bottom: Collapse of the scaled excess penetration into vacuum for different times in both the AP and the REP. Inset: The scaled excess penetration decays as a (3/2)-stretched exponential in both models. Data for the REP have been shifted vertically for clarity.}
\label{profs}
\end{figure}

To estimate $\mathcal{P}(\eta)$ in the opposite $\eta\to \infty$ limit, it is convenient to consider a discrete-time version of the AP model with $\rho_0=1$, namely we pick up a pair of neighbors and make a collision with probability equal to the pair energy. We want to transfer energy to the right as far as possible. To this end we select the front pair $(\ell_+,1+\ell_+)$  at each step $j=\ell_+\geq 0$ for a collision, even though the pair collision rate is decaying as $2^{-j}$, i.e. as the energy at the rightmost site decreases. The front moves ballistically, $\ell_+(t)=t$, which implies $\eta\sim t^{1-\mu}$ via \eqref{j+}, and the logarithm of the total probability of the above process scales as $\ln \prod_{j=0}^{t-1}2^{-j}\sim -t^2 \sim -\eta^{2/(1-\mu)}$. Combining both asymptotics, we arrive at
\begin{equation}
\label{tails}
\ln \mathcal{P}(\eta) \sim
\begin{cases}
- |\eta|^{2/(1-3\mu)}    & \eta\to -\infty\\
- \eta^{2/(1-\mu)}        & \eta\to  \infty
\end{cases}
\end{equation}
These estimates rely on the validity of Eq.~\eqref{j+} for large values of the random fluctuation where $\eta$ diverges as a power of time. This assumption is known to be valid in numerous examples \cite{book}, where matching to the extreme typically allows to determine the tail scaling.

We measured $\mathcal{P}(\eta)$ for the AP and REP models, see inset to Fig.~\ref{front}. Data are consistent with a faster decay of negative front fluctuations for the REP as compared to the AP ($\ln \mathcal{P}(\eta) \sim - |\eta|^4$ vs $- |\eta|^2$ as $\eta\to -\infty$, respectively), although our extensive simulations did not allow us to access the rare-event tails where the asymptotic behavior \eqref{tails} becomes clear. The probability of positive fluctuations is closer for the REP and the AP, as expected. Novel Monte Carlo methods allowing to probe rare events may help in measuring these tails \cite{kurchan,pablo}.

Top panel in Fig. \ref{profs} shows the scaled energy profile measured for different times in AP model, together with $\Phi(\xi)$ of Eq.~\eqref{parabola}. The agreement is excellent for intermediate and long times ($t>10^2$). Moreover, fluctuating corrections to the hydrodynamic profile can be measured, see top inset in Fig. \ref{profs}, revealing a microscopic penetration into vacuum of energy profiles which exhibits scaling behavior and robust features. We define the excess penetration into vacuum as $\Delta \Phi\equiv t^{1/3}\la\rho(x,t)\ra-\Phi(xt^{-1/3})$. Data for different times nicely collapse onto a single curve under the scaling $\Delta\Phi(\xi)=t^{-(1-3\mu)/3}\mathcal{F}[(\xi-\xi_0)t^{(1-3\mu)/3}]$, see bottom panel in Fig. \ref{profs}, with $\mathcal{F}(z)$ a scaling function slightly different for the AP and REP models. Despite this difference, bottom inset to Fig. \ref{profs} shows that excess penetration into vacuum decays as a (3/2)-stretched exponential in both models, i.e. $\mathcal{F}(z)\sim \exp[-(z/\chi)^{3/2}]$ for $z>0$, where $\chi$ is a scaled penetration length.

\begin{figure}
\vspace{-0.5cm}
\centerline{
\vspace{-0.25cm}
\includegraphics[width=7cm]{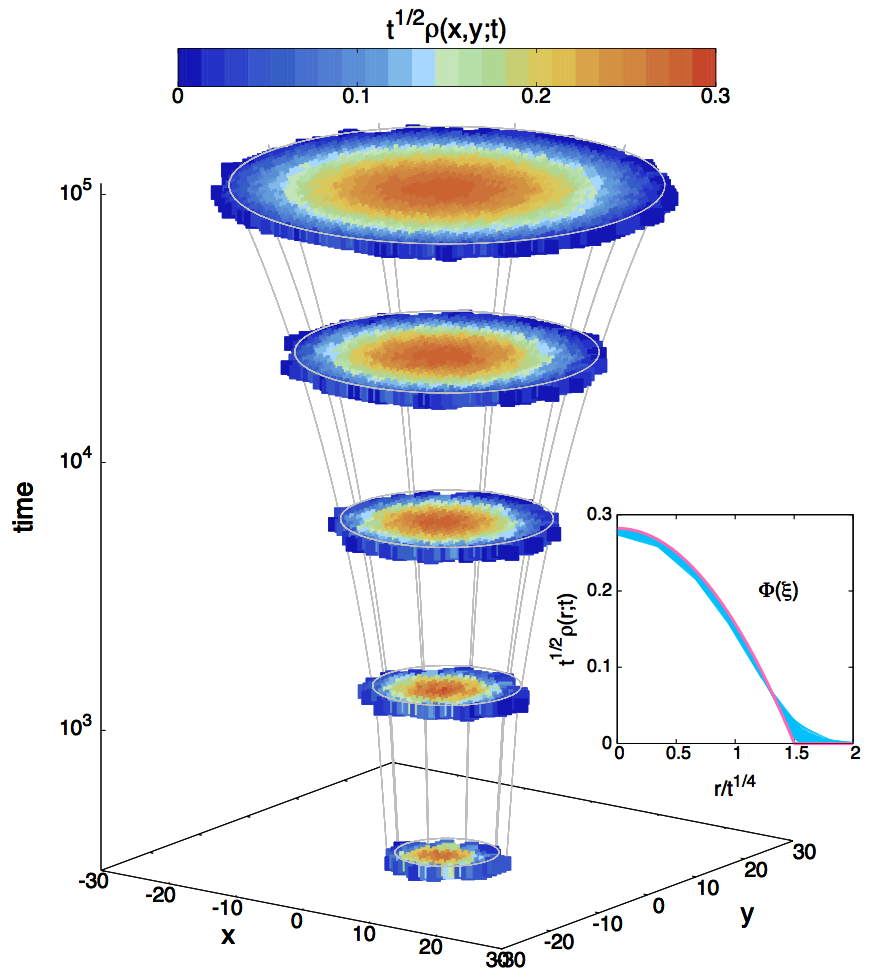}}
\caption{(Color online) Space-time evolution of the scaled energy for the 2d AP model. Inset: Convergence of the scaled radial profile for different times to the hydrodynamic solution.}
\label{twodim}
\end{figure}

The generalization of the AP and the REP models to higher dimensions is straightforward. In this case the governing hydrodynamic equation is equivalent to \eqref{NLDE:general} with $a=1$. A localized energy pulse evolves again into a self-similar wave with compact support characterized by a parabolic scaled profile similar to \eqref{parabola}, see Fig.~\ref{twodim}, whose front advances as $r_f(t)=\xi_0 (\rho_0 t)^{1/4}+(\rho_0 t)^{\mu}\eta$. Front fluctuations in 2d are strongly suppressed: $\mu\approx 1/30$ for the REP, and again $\mu=0$ for the AP. In 2d we also observe a similar excess penetration into vacuum of energy profiles, characterized by a scaling function with (3/2)-stretched exponential decay as in 1d, and a scaled penetration length $\chi$ which decreases with $d$. This observation, together with the small value of exponent $\mu$, suggests that fluctuating corrections to hydrodynamic behavior become less pronounced in high dimensions.

\begin{figure}
\vspace{-0.5cm}
\centerline{
\vspace{-0.5cm}
\includegraphics[width=8cm]{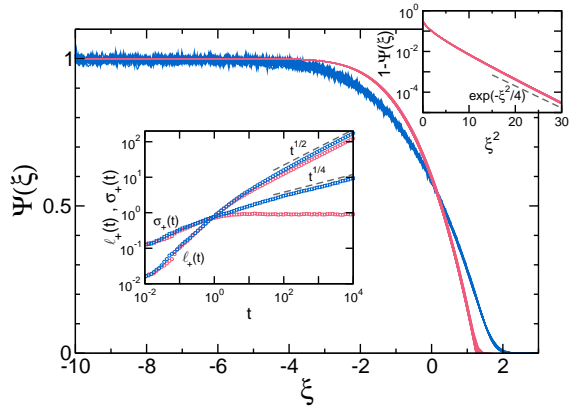}}
\caption{(Color online) Scaled rarefaction wave profile for the AP (red) and the REP (blue) models. Top inset: Asymptotic profile behavior for $\xi\to -\infty$ in the AP. Bottom inset: Front position ($\bigcirc$) and its standard deviation ($\Box$).}
\label{step}
\end{figure}

We return now to 1d and consider the discharge of a step initial profile into vacuum \cite{ASEP}. Such initial condition evolves into a self-similar, compact rarefaction wave. In the hydrodynamic approximation we need to solve Eq.~\eqref{NLDE}  subject to $\rho(x,t=0)=1$ for $x<0$ and $\rho(x,t=0)=0$ for $x>0$. The transformation $x\to  \sqrt{\beta}\, x, ~ t\to \beta t, ~ \rho\to \rho$ leaves the 
initial-value problem invariant.  The relevant self-similar solution is therefore $\rho(x,t)=\Psi(\xi)$ with $\xi = x/\sqrt{t}$, and the scaling function obeys $\Psi \Psi'' + (\Psi')^2 + \tfrac{1}{2}\xi\Psi' = 0$. This equation does not admit an analytical solution, though it is still possible to understand the qualitative behavior of $\Psi(\xi)$. The initial condition implies that $\Psi\to 1$ when $\xi\to -\infty$ and $\Psi\to 0$ when $\xi\to\infty$. An assumption that the solution spreads into the vacuum ad infinitum leads to a contradiction \cite{longpaper}. Hence similarly to the compact solution \eqref{parabola}, the energy profile vanishes at a finite value of the scaling variable $\xi=\xi_0$, see Fig. \ref{step}. An asymptotic analysis in the proximity of the front shows that $\Psi\simeq \tfrac{1}{2}\xi_0(\xi_0-\xi)$ as $\xi\to \xi_0$. In the opposite $\xi\to -\infty$ limit we employ a perturbative analysis to yield $1-\Psi\sim  e^{-\xi^2/4}$  \cite{longpaper}. This asymptotic behavior, confirmed in the top inset to Fig. \ref{step}, is not surprising since in the far-left region the governing equation is essentially a linear diffusion equation. Despite the overall diffusive scaling, there is a well-defined front separating the perturbed region from the vacuum, which propagates in time as $\ell_+(t) = \xi_0\sqrt{t} + t^\delta \eta$ with $\eta$ again a random variable. Bottom inset in Fig. \ref{step} shows that $\delta=1/4$ for the REP and $\delta=0$ for the AP. We can analyze in equivalent terms the tails of the distribution $\mathcal{P}(\eta)$ of front fluctuations \cite{longpaper,ASEP}. Furthermore, the scaled rarefaction waves in Fig. \ref{step} also exhibit microscopic penetration into vacuum, characterized by scaling functions $\mathcal{F}(z)$ with $(3/2)$-stretched exponential tail as in Fig.~\ref{profs}.

Our results suggest the existence of robust scaling behaviors at the fluctuating level in nonlinear diffusion models. This is remarkable since, despite the macroscopic similarity between the AP and REP models, there is a crucial difference --- the REP is time-reversible (i.e., it obeys the detailed balance condition), while the AP lacks microscopic reversibility as collisions cannot be reversed. The observed signs of universality are thus intriguing and deserve further study. The class of microscopic diffusive models introduced here represent an ideal benchmark to investigate strongly nonlinear phenomena far from equilibrium, and can be generalized to describe nonlinear driven dissipative media \cite{plh}. 
%In addition to the AP and REP models, the Brownian energy process \cite{BEP} also yields generalized nonlinear diffusion in certain limits. 
The simplicity of these models allows to go beyond hydrodynamics to study fluctuations in nonlinear, non-equilibrium settings. This is particularly interesting at the light of the recent breakthroughs in non-equilibrium physics within the Hydrodynamic Fluctuation Theory formalism \cite{pablo,bertini}. In fact, it would be interesting to study rare fluctuations and large-deviation properties for the nonlinear diffusive models introduced here, in particular in a dynamical setting as the compact wave propagation problem \cite{longpaper}, thus complementing recent results on front and current fluctuations for shocks in simple exclusion processes \cite{ASEP}.

PIH thanks J.~L.~Lebowitz for useful discussions. Financial support from MICINN project FIS2009-08451, University of Granada, Junta de Andaluc\'{\i}a projects P07-FQM02725 and P09-FQM4682  is acknowledged.

\end{document}